\pdfoutput=1

\documentclass[12pt,draftclsnofoot, onecolumn]{IEEEtran}
\usepackage{amsmath,algorithm,algpseudocode}
\usepackage{amssymb}
\usepackage{array}
\usepackage{times,epsfig,soul,subfigure,color}

\begin{document}
\pagenumbering{gobble}

\title{Learning Sequential Channel Selection for Interference Alignment using Reconfigurable Antennas}

\author{
Nikhil Gulati$^{\dagger}$, Rohit Bahl$^{\ddagger}$, and Kapil R. Dandekar$^{\dagger}$ %
\vspace{12pt}\\
$^{\dagger}$Department of Electrical and Computer Engineering, Drexel University,\\
$^{\ddagger}$ Intel Mobile Communications\\

Email: ng54@drexel.edu, rohit.bahl@intel.com, dandekar@coe.drexel.edu 
}

\maketitle

\begin{abstract}
In recent years, machine learning techniques have been explored to support, enhance or augment wireless systems especially at the physical layer of the protocol stack. Traditional ML based approach or optimization is often not suitable due to algorithmic complexity, reliance on existing training data and/or due to distributed setting. In this paper, we formulate a reconfigurable antenna based channel selection problem for interference alignment in a multi-user wireless network as a learning problem. More specifically, we propose that by using sequential learning, an effective channel or combination of channels can be selected in order to enhance interference alignment using reconfigurable antennas. We first formulate the channel selection as a multi-armed problem that aims to optimize the sum rate of the network. We show that by using an adaptive sequential learning policy, each node in the network can learn to select optimal channels without requiring full and instantaneous CSI for all the available antenna states. We conduct performance analysis of our technique for a MIMO interference channel using a conventional IA scheme and quantify the benefits of pattern diversity and learning channel selection.
\end{abstract}



\section{Introduction}
\label{intro}
As the number of wireless network users grow exponentially, the network becomes denser and interference becomes a serious bottleneck. Designing robust interference management techniques for future multiuser wireless networks is of great importance for enhancing network capacity and supporting multiple users. Further, the dynamic nature of wireless channel, distributed users and environment make it very difficult to optimize any interference management techniques for all environments. Applying machine learning techniques to such problems may provide viable solutions. But tradtional supervised techniques or deep learning techniques require a large amount of training data and also have significant computational complexity. On the other hand reinforcement learning or sequential learning allows more natural setup for communication problems to learn optimization functions in an unknown environment and applications such as spectrum sensing, signal detection and interference management. 

Among various interference management techniques, Interference Alignment (IA) has generated a lot of work and excitement. Initially proposed in~\cite{maddah2008communication}, the authors in~\cite{cadambe2008interference} made an important advancement in this direction by utilizing Interference Alignment (IA) and proving that the sum capacity of a {multiuser} network is not fundamentally limited by the amount of interference. Subsequently, they showed IA based precoding to achieve linear scaling of Degrees of Freedom (DoF) and sum capacity in the high signal-to-noise ratio (SNR) regime.


The key insight for IA is that perfect signal recovery is possible if interference does not span the entire received signal space. As a result, a smaller subspace free of interference can be found where the desired signal can be projected while suppressing the interference to zero. For perfect IA, where all the transmitters and receivers have perfect channel state information (CSI) available, the requirement is to design the transmit precoding and receive decoding filters to maximize the interference free dimensions. It has been shown that designing such filters is generally NP hard~\cite{razaviyayn2010linear} except for a few cases such as shown in~\cite{cadambe2008interference} and~\cite{tresch2009achievability}. As an alternative to finding closed-form solutions, many algorithmic techniques have been proposed in the literature for IA~\cite{gomadam2011distributed},~\cite{peters2009interference},~\cite{yu2010least}. The basic idea is to minimize leakage interference at each receiver to achieve perfect alignment for the best case scenario.


Further, even if the perfect alignment is achieved, it does not guarantee maximum sum capacity in linear receivers due to loss of SNR. This loss of SNR prevents IA solutions from being optimal in low or mid SNR regimes for practical implementation. Since the component of the desired signal lying in the interference space is lost after projection, the sum capacity scaling achieved comes at the expense of reduced SNR~\cite{sung2010linear}. Therefore, the key insight from~\cite{sung2010linear} is that in order to achieve optimal performance, the two spaces must be roughly orthogonal.

In~\cite{el2010feasibility}, the authors show that orthogonality of the subspaces is influenced by the nature of the wireless channel and hence may not always be achievable in the real world. Further, the authors provided a feasibility study of IA over measured channels and established an empirical relation between sum capacity and distance between the signal and interference space. They quantified the effect of correlated channels on the sum capacity and showed the sub-optimality of IA at low SNR. Another experimental study reported in~\cite{gonzalez2011experimental} showed similar degradation in the performance of IA because of practical effects such as collinearity of subspaces arising in real world channels. An alternative approach to improve orthogonality between the subspaces in order to maximize sum capacity was shown~\cite{6477634},~\cite{el2012improved}~\cite{6594784} using the pattern diversity of reconfigurable antennas. Reconfigurable antennas are capable of dynamically altering their radiation characteristics in response to the needs of the underlying network. In~\cite{6477634}, the authors used multiple states of the reconfigurable antenna for selecting optimal channel coefficients using exhaustive search and experimentally showed using measured channels, the increase in sum capacity over conventional IA even in low SNR regimes. The authors in~\cite{6594784}, proposed a set of sequential algorithms to select the channel coefficient with and without perfect CSI and quantified the BER performance along with analysis for outage probability. The proposed work in this paper extends this concept and we propose a sequential learning algorithm for selecting optimal channel coefficients for IA precoder and decoder design.

Traditionally, translating the benefits of reconfigurable antennas into a practical wireless systems is a challenging task~\cite{gulati2014learning},~\cite{li2012algorithms}. Both in single and multi user wireless networks, the cost of CSI acquisition and channel training for all the states of the reconfigurable antenna can sometimes negate the diversity benefits. Therefore, novel techniques to amortize that cost are important for successful and widespread integration of reconfigurable antennas in future wireless systems and standards.

Towards addressing these challenges, in this paper, we first formulate the sequential channel selection as a multi-armed bandit problem and show how each node can sequentially explore the channel vector space to maximize a network reward function. Our proposed technique is different than the sequential algorithm in~\cite{6594784} which follows a fixed switching structure (static) and relies on heuristics to find the optimal channel coefficients. Further, their scheme does not scale with the number of antenna states. Our learning technique is dynamic in nature and adapts the exploration process based on the observed interference and desired channel coefficients. We provide analysis in terms of the improvements achieved in sum capacity, and the distance between interference and desired signal space. We also provide empirical analysis of the cost of learning the optimal channel coefficients to provide some bounds on the performance and convergence of the algorithms. 

Our proposed learning technique is dynamic in nature and adapts the exploration process based on the observed interference and desired channel coefficients. We provide analysis in terms of the improvements achieved in sum capacity, and the distance between interference and desired signal space. We also provide empirical analysis of the cost of learning the optimal channel coefficients and convergence of the algorithms. 



Specifically, we provide the following contributions in this work:

\begin{enumerate}
	\item We first formulate the reconfigurable antenna based channel selection problem under the multi-armed bandit framework taking into account the distributed network setting of IA.
	\item We identify key metrics which can be used with the learning framework to assess the long-term performance of the system. 
	\item We present sequential learning algorithms in both centralized and distributed setting to provide the first practical technique to integrate reconfigurable antennas in systems making use of IA.
\end{enumerate}

The rest of the paper is organized as follows: In Section~\ref{relatedwork}, we provide a background on reconfigurable antenna applications in wireless networks and multi-armed bandit theory along with related work on both topics. Section~\ref{systemModel} describes a system model for K-user MIMO interference channel for IA and employing reconfigurable antennas. We also describe the multi-armed bandit formulation for sequential channel selection in Section~\ref{banditform}. In Section~\ref{policydes}, we describe the selection policies and the reward metrics used to evaluate the performance of the proposed schemes. In Section~\ref{perfAnalysis}, we provide a description of simulation setup, evaluated algorithms and performance analysis, followed by the conclusion in Section~\ref{conclusion}.

\section{Background and Related Work}
\label{relatedwork}
\subsection{\textbf{Channel Selection for Reconfigurable Antennas}}
By providing multiple, potentially uncorrelated, channel realizations, reconfigurable antennas provide additional degrees of freedom for adaptation~\cite{piazza2009two} as well as provide space-cost benefits~\cite{sukumar2009link}.  These antennas have been shown to enhance the performance of single user MIMO systems by increasing the channel capacity, diversity order~\cite{boerman2008performance},~\cite{piazza2008design} and even have been shown to perform well in low SNR regimes,~\cite{sayeed2007maximizing}. They have also been to be beneficial for new wireless applications such as physical layer security~\cite{gulati2013gmm}, key generation~\cite{6834803} and spectrum sensing~\cite{wanuga2014online}. For reaping maximum benefits, a channel (alternatively antenna state) selection technique to identify optimal channel coefficients is required. 


Periodic exhaustive training techniques with reduced overhead is presented in~\cite{eslami2010reduced} where the authors highlight the effect of channel training frequency on the capacity and the bit-error rate (BER) of a MIMO system. More recently, in~\cite{gulati2014learning}, the authors proposed an online learning based framework for antenna state selection in single user MIMO systems with experimental validation. More closely related to the work presented in this paper, in~\cite{6477634},~\cite{el2012improved}~\cite{6594784}, the authors have proposed to use the multiple states of a reconfigurable antenna to enhance the performance of IA. More specifically, in~\cite{6594784}, the authors proposed two sequential antenna state switching techniques corresponding to closed-form IA and channel reciprocity based distributed IA schemes. Further, the authors show the SINR and BER improvements of IA by switching to optimal antenna states. The authors further determine that as the number of users and number of reconfigurable antenna states increase, the problem becomes NP-hard and can only be solved using approximate algorithms or stochastic optimization. They restrict the number of antenna states to two in order to apply heuristic or brute force search techniques to find the best candidate solution. In this paper, we neither assume full CSI for all antenna states at every time slot nor do we perform periodic exhaustive search. Our work (See Sec.~\ref{banditform}) relies on an approximate solution via online learning to learn and track the derived channel metrics which adapts the state selection policy accordingly. In other words, our proposed technique can scale with the number of antenna states and allows for lower computationally complexity in order to find the best antenna state combinations at the receiver.

\subsection{\textbf{Multi-Armed Bandit (MAB) for Wireless Networks}}
Multi-armed bandit theory provides a mathematical framework to learn unknown parameters of a distribution via online learning\cite{vermorel2005multi,Lai19854}. It represents the well known exploitation vs exploration dilemma in environments with partial or no information.  Application of online learning has also been investigated in network optimization~\cite{gai2012combinatorial}, as well as opportunistic and dynamic spectrum access~\cite{liu2010distributed}. The bandit formulation is applied to the cellular coverage optimization in~\cite{shen2018generalized}. Further in~\cite{gai2010learning}, a combinatorial version of MAB was proposed for a network with multiple primary and secondary users taking into account the collisions among the secondary users. Distributed channel allocation among multiple secondary users was further studied and proposed in~\cite{gai2011decentralized} and~\cite{liu2010distributed}. Another application of multi-armed bandit for cognitive radio was proposed in~\cite{volos2010cognitive} for adaptive modulation and coding. More recently, authors in~\cite{amuru2016jamming} have shown the use of multi-armed bandit framework to design optimal startegies for a cognitive wireless jammer. 

 
\textsl{Notation:} We use capital bold letters to denote matrices and small bold letters for vectors. $\textbf{H}^{-1}$, $\textbf{H}^{\dagger}$ and $\textbf{H}^{T}$ denote the matrix inverse, Hermitian and transpose operation respectively. Span(\textbf{H}), null(\textbf{H}) and $\left\|\textbf{H}\right\|_{F}$would represent the space spanned by the columns of \textbf{H}, the null space of \textbf{H} and Frobenius norm of $\textbf{H}$ respectively. The d $\times$ d identity matrix is represented by $\textbf{I}_{d}$.
\section{System Model for Interference Alignment using Reconfigurable Antennas}
\label{systemModel}
Consider the $K$ user MIMO interference channel in which each transmitter (Tx) is equipped with $M$ conventional omni-directional antennas and each receiver (Rx) is equipped with $N$ reconfigurable antennas. The reconfigurable antennas at the receiver have $\mathcal{P}$  reconfigurable states from which to choose. Each of these states correspond to a unique radiation pattern. In such a setting, the received signal at the $i^{th}$ receiver can then be represented by
\begin{equation}
\label{signal}
\textbf{y}^{\left[i\right]}=\textbf{H}^{\left[i,i\right]}_{p}\textbf{x}^{\left[i\right]} + \sum_{\substack{k=1 \\ k \neq i}}^{K}{\textbf{H}^{\left[i,k\right]}_{p}}\textbf{x}^{\left[k\right]} + \textbf{n}^{\left[i\right]},
\end{equation}
where $p$ represents the antenna state selected at the receiver, $\textbf{y}^{\left[i\right]}$ is the $N \times 1$ received column vector, $\textbf{H}^{\left[i,k\right]}_{p}$ is the $N \times M$ MIMO channel between Tx $k$ and Rx $i$, $\textbf{x}^{\left[k\right]}$ is the $M \times 1$ input column vector and $\textbf{n}$ represents the $N \times 1$ vector of complex zero mean Gaussian noise. Also, $\textbf{H}^{\left[i,k\right]}_{p}$=$\textbf{H}^{\left[i,k\right]}\mathbf{R_{r}}$ where $R_{r}$ is an $N \times M$ diagonal matrix with $n^{th}$ diagonal entry $\sigma^{2}_{p,n}$, which defines the mean power of each state $p$. 


Further, $\textbf{H}^{\left[i,i\right]}_{p}$ is generated in the same way. We note that the off-diagonal elements of $R_{r}$ are zero since we do not consider correlated channels. 
The achievability of MIMO IA for correlated channels is still not fully understood and has been studied only for transmit antenna correlation~\cite{nosrat2011mimo}.


Throughout this paper, we will restrict our study to $K = 3$; $M=N=2$ and $d_{k} = 1$, $\forall$ $k$ $\in$ $\{1,2,3\}$.


\subsection{Interference Alignment for the 3 user, 2$\times $2 MIMO Channel}
\label{IA3users}
The goal of IA is to make the signal to interference ratio (SIR) infinite at the output of each receiver by designing precoders and decoding filters to eliminate interference. 


It has been shown that designing such precoding filters is NP hard in general for MIMO systems~\cite{razaviyayn2010linear} and closed form solutions exist only for certain special cases such as the three user $2\times2$ MIMO channel~\cite{cadambe2008interference}. More practical approach to achieve IA was proposed in~\cite{gomadam2011distributed}.


Let $\textbf{v}^{\left[i\right]}$ and $\textbf{u}^{\left[i\right]}$ represent the transmit precoder and receive interference suppression filter respectively, where $i$ $\in$ $\{1,2,3\}$ and $\textbf{v}^{\left[i\right]}$, $\textbf{u}^{\left[i\right]}$ $\in$ $\mathbb{C}^{2\times1}$. 


After precoding the input symbol ${x}^{\left[i\right]}$ with $\textbf{v}^{\left[i\right]}$, the signal received at the $i^{th}$ receiver can be represented by (\ref{sig}). 


\begin{gather}
\label{sig}
\textbf{y}^{\left[i\right]} = \textbf{H}^{\left[i,i\right]}\textbf{v}^{\left[i\right]}{x}^{\left[i\right]} + \sum_{\substack{k=1 \\ k \neq i}}^{K}{\textbf{H}^{\left[i,k\right]}}\textbf{v}^{\left[k\right]}{x}^{\left[k\right]} + \textbf{n}^{\left[i\right]} \\
\label{span1}
\end{gather}

When the interference is completely eliminated as a result of alignment, the interference $\textbf{H}^{\left[i,j\right]}\textbf{v}^{\left[j\right]}, j\in\{1,2,3\}, j\neq i$, is aligned in the same subspace (direction) and different from the subspace (direction) of the desired signal $\textbf{H}^{\left[i,i\right]}\textbf{v}^{\left[i\right]}$. 

The angle between the effective interference and desired channels as given above can be changed by manipulating the corresponding $\textbf{H}$. With the availability of reconfigurable antennas at the wireless node, this diversity in $\textbf{H}$ can be exploited to select effective channel vectors in both interference and signal subspace to maximize the angle between the two sub spaces. The authors in~\cite{6594784} analytically proved that the SINR at any receiver $k$ is maximized when the two subspaces are orthogonal. When the number of antennas is $N=2$ at each receiver, all the vectors $\textbf{u}^{\left[i\right]\dagger}$, $\textbf{H}^{\left[i,j\right]}\textbf{v}^{\left[j\right]}$ and $\textbf{H}^{\left[i,i\right]}\textbf{v}^{\left[i\right]}$ are two dimensional vectors lying in the same plane.  Therefore, for the case of $N=2$, the SINR is maximized at $\theta=\frac{\pi}{2}$ where $\theta$ is the angle between the subspaces. We will show later that instead of measuring the angle between the subspaces, it is more effective to measure the subspace distance defined over a grasmman manifold~\cite{love2005limited}.


\subsection{Mulit-Armed Bandit Formulation for Sequential Channel Selection}
\label{banditform}


We first consider a system with $\mathcal{P}$ unknown random processes, $X_{i}(n), 1\leq i \leq \mathcal{P}$ where $n$ is used to index discrete time steps. We also assume that $X_{i}(n)$ evolves as an i.i.d random process with an arbitrary distribution with finite support. Without loss of generality, we can normalize $X_{i}(n)\in \left[0,1\right]$. We denote the expected value of this random process with mean $\mu_{i}$ which is unknown to the users.

At each time slot $n$, the bandit controller selects a $\mathcal{K}$-dimensional combinational vector of coefficients $c(n)$ from a finite set of $I$. We assume that all the elements $c_{i}^{j}(n) > 0$ for all $1 \leq i \leq \mathcal{P}$ and $1 \leq j \leq K$. Further, the reward obtained from selecting a combinational vector $c$ is given by:

\begin{equation}
\label{totReward}
R_{c(n)}(n)=\sum_{j=1}^{K}X_{i}(n)
\end{equation}

We now map the described formulation to $K$-user interference network described in section~\ref{systemModel}. In this set up, $K$ wireless receivers employ pattern reconfigurable antennas. The receivers can select from $\mathcal{P}$ available antenna states which reduces the problem to selecting a $K$-dimensional combinational vector of antenna states where each receiver can select a radiation state $i$ independently from $\mathcal{P}$. The decision is made at every time slot (packet) $n$, to select the combinational vector of antenna states to be used for the next reception.

In practice, this is performed post alignment, where the best beamforming vectors and decoders are selected given the channel associated with each combination $c$. When each receiver selects an available antenna state $i$, an instantaneous random reward is achieved, which we denote as $X_{i}\left(n\right)$. Next, the bandit controller calculates the total reward given by Eq.~\ref{totReward}. Also, the reward is only observed for the selected state combination $c$ and not for the other state combinations. In other words, the bandit controller receives channel state information for only the selected radiation state combination $c$ at a given time slot and acquires no new information about the other possible combination state vectors. In this way, our proposed technique differs from the other techniques in the literature as it does not rely on the availability of instantaneous CSI for all the radiation state combinations available for each receiver, at each time slot.



The goal of the multi-armed bandit policies is to perform well with respect to \textit{regret}.  Regret is defined as the difference between the expected reward that can be obtained by an oracle which always picks the optimal action at each time slot (or through exhaustive search) and the reward obtained by a given multi-armed bandit algorithm. This difference is also sometimes referred to as sub optimality gap. The goal of the bandit policies is to minimize regret, or in other words, maximize expected reward. Regret for a policy can then be calculated as

\begin{equation}
\label{laiRegret}
\mathcal{R}_{n}^{\pi}=n\mu^{*}-E^{\pi}[\sum_{i=1}^{n}R_{\pi(t)}(t)]
\end{equation}
where,
\begin{equation}
\mu^{*}=\text{max}_{c} \sum_{c=1}^{I}R_{c}
\end{equation}

$\mu^{*}$ is the expected reward of the optimal combinational vector of antenna states. 

Intuitively, the regret $\mathcal{R}_{n}^{\pi}$ should be as small as possible. Well-performing multi-armed policies aim to achieve sublinear regret w.r.t to time $n$ which can result in time-averaged regret to tend to zero and achieving maximum possibly regret. It has been shown in~\cite{Lai19854} that the minimum rate at which regret grows is of logarithmic order under certain regularity conditions. The authors established that for some families of reward distributions there are policies that can satisfy

\begin{equation}
\label{satcond}
E\left[T_{i}\left(n\right)\right]\leq\left(\frac{1}{D\left(\mu_{i}||\mu^{*}\right)}+o(1)\right)\text{ln}(n)
\end{equation}
where $o\left(1\right)\rightarrow 0$ as $n\rightarrow \infty$ and 

\begin{equation}
\label{kldivergence}
D\left(\mu||\mu^{*}\right) \equiv\int \mu_{i} \text{ln} \frac{\mu_{i}}{\mu_{*}}
\end{equation}

is the Kullback-Leibler divergence between the reward density $\mu_{i}$ of a suboptimal arm $i$ and the reward density of the optimal arm $\mu^{*}$. Over an infinite horizon, the optimal arm is expected to be played exponentially more often than any other arm.

\section{Policy Design and Reward Metrics}
\label{policydes}

\subsection{Sequential Policy Design}

In this section we will describe the criteria for choosing the multi-armed bandit learning policies. We consider a class of non-Bayesian non-parametric multi-armed bandit algorithms which work by associating an index called \textit{upper confidence index} to each arm. The calculation of such an index relies on the entire sequence of rewards obtained up to a point (time slot) from selecting a given arm. The computed index for each arm is an estimate for the corresponding reward expectations.

We first show the most commonly used index policy for the non-Bayesian case, known as the UCB1 policy~\cite{auer2002finite}. We will then explore a more robust and flexible index based policy known as KL-UCB~\cite{garivier2011kl} and will show a modified algorithm for this channel selection problem (described in Sec.~\ref{banditform}) in section~\ref{rewardMetrics}.

\subsubsection{UCB1 - Selection Policy} To implement the UCB1 policy, the bandit controller stores two variables. The first variable is the average reward $\bar{R}_{c}(n)=\sum_{i=1}^{n}R_{c(n)(i)}$ up to the current packet $n$, where $R_{c(n)}(n)$ is defined in Eq.~\ref{totReward}. Further, the bandit controller stores the number of times the combination vector $c$ has been selected up to the current packet $n$, denoted by ${m}_{c}\left(n\right)$. 

The UCB1 policy as shown in algorithm~\ref{alg1}, first begins by selecting each antenna state combination vector $c$ at least once and $\bar{R}_{c}\left(n\right)$ and ${m}_{c}\left(n\right)$ are then updated using(~\ref{rewardUpdateRule}) and(~\ref{numplaysUpdateRule}).

\begin{equation}
\label{rewardUpdateRule}
\bar{R}_{c}(n) =\left\{\begin{array}{l l} \frac{\bar{R}_{c}\left(n-1\right){m}_{c}\left(n-1\right) + R_{c(n)}\left(n\right)}{{m}_{c}\left(n-1\right)+1} & \quad \text{if \textit{c} is selected}\\\\
\bar{R}_{c}\left(n-1\right) & \quad \text{else}\\\\
\end{array} \right.
\end{equation}

\begin{equation}
\label{numplaysUpdateRule}
{m}_{c}\left(n\right) =\left\{\begin{array}{l l} {m}_{c}\left(n-1\right)+1 & \quad \text{if \textit{c} is selected}\\
{m}_{c}\left(n-1\right) & \quad \text{else}\\
\end{array} \right.
\end{equation}

As shown in algorithm~\ref{alg1}, once the initialization is completed, the policy selects the combination vector that maximizes the criteria on line~\ref{UCB1maxCriteria}. From line~\ref{UCB1maxCriteria}, it can be seen that the index of the policy is the sum of two terms. The first term is simply the current estimated average reward (sample mean) for a given combination vector. The second term is the size of the one-sided confidence interval of the estimated average reward within which the true expected value of the mean falls with a very high probability. As an antenna state combination is selected more often, the estimate of the average reward improves and the confidence interval size reduces. Eventually, the estimated average reward reaches as close as possible to the true mean. The size of the confidence interval also governs the index of the arm for future exploration. 



\begin{algorithm}[ht]
\caption{UCB1 Policy, Auer et al.~\cite{auer2002finite} }
\label{alg1}
\begin{algorithmic}[1]
\State // Initialization
\State ${m}_{c} , \bar{R}_{c} \leftarrow 0$
\State Select each state combination $c$ at least once and update ${m}_{c} , \bar{R}_{c}$ accordingly.
\State // Main Loop
\While {$1$}
\State	Select $c$ that maximizes $\bar{R}_{c} + \sqrt{\frac{2\text{ln}(n)}{{m}_{c}}}$ \label{UCB1maxCriteria}
\State  Update ${m}_{c}$, $\bar{R}_{c}$ for the selected state combination $c$
\EndWhile
\end{algorithmic}
\end{algorithm}

In terms of regret, the UCB1 policy has an expected regret of at most~\cite{auer2002finite}:

\begin{equation}
\label{UCB1maxregret}
\left[8 \sum_{i:\mu_{i}<\mu^{*}}\frac{\text{ln}n}{\Delta_{i}}\right]+\left(1+ \frac{\pi^{2}}{3}\right)\left(\sum_{i:\mu_{i}<\mu^{*}}\Delta_{i}\right)
\end{equation}

where $\Delta_{i}=\mu^{*}-\mu_{i}$

We have shown through our previous work~\cite{gulati2014learning}, that UCB and its variants work well for link throughput optimization in a single user MIMO systems by using appropriate reward functions. In this work we apply a more robust and stable learning policy known as KL-UCB~\cite{garivier2011kl}

KL-UCB has been shown to provide strictly better theoretical regret guarantees than UCB while maintaining the same applicability~\cite{maillard2011finite}. Further, what makes it more suitable for this work is the ability to modify the algorithm for arbitrary reward distributions while maintaining regret guarantees.

The KL-UCB policy stores three variables to find the optimal arm both in the finite horizon as well as infinite horizon setting. Therefore, in this case the bandit controller will store a) the number of times an arm is used b) the total reward an arm has received and c) a real non-negative parameter $a$, generally set to zero.

For the channel selection problem described above, an arm represents the combination vector $c$ of the antenna states selected at each receiver, total reward as defined in Eq.~\ref{totReward} is the sum of the rewards received from each receiver up to $n$. 

The KL-UCB policy as shown in algorithm~\ref{alg2}, starts by playing each arm once to initialize it. Then, once all the arms are initialized, we calculate the upper confidence bound on line~\ref{KLUCBmaxCriteria} using the divergence of the form, $d(x,y)=\frac{x}{y}-1-log(\frac{x}{y})$. The arm with the highest index is selected for the next round and the ${m}_{c}$, $\bar{R}_{c}$ are updated accordingly.

\begin{algorithm}[ht]
\caption{KL-UCB Policy, Cappe et al.~\cite{garivier2011kl} }
\label{alg2}
\begin{algorithmic}[1]
\State // Initialization
\State ${m}_{c} , \bar{R}_{c} \leftarrow 0$
\State Select each state combination at least once and update ${m}_{c} , \bar{R}_{c}$ accordingly.
\State // Main Loop
\While {$1$}
\State $c \leftarrow \text{argmax}_{c \in I}\text{max}[m_{c}d(\frac{\bar{R}_{c}}{m_{c},1})\leq log(t)]$ \label{KLUCBmaxCriteria}
\State  Update ${m}_{c}$, $\bar{R}_{c}$ for state combination $c$
\EndWhile
\end{algorithmic}
\end{algorithm}

We note that while using both UCB1 and KL-UCB, any combination vector $c$ which denotes the arm is a vector of individual antenna states at each receiver. In that case, we ignore any direct correlation between the antenna states arising due to any pattern correlation for the purpose of this work. We believe this could be an interesting area of future work where a bandit policy is constructed to take into account any correlation between the coefficients of a combination vector~\cite{gai2012combinatorial}.

\subsection{Reward Metrics}
\label{rewardMetrics}

In this section, we discuss the metrics derived at the receiver post alignment to use as rewards for the bandit formulation described above. The selection of reward metrics is dependent on the specific system implementation and based on the desired objective, the system designer can identify a relevant reward metric. In this paper, we evaluate two commonly used network performance metrics for interference limited wireless networks. 

\subsubsection{Network Sum Rate}
In most existing research in IA, network sum rate is used as the key indicator for measuring the performance of any IA algorithm. Thus, we use sum rate as the first reward function to be used with the bandit policies described above. With successful alignment, the rate achieved by the $k$th user can be defined as


\begin{equation}
\label{rate}
R^{[k]}=\text{log}\left|\textbf{I}_{d^{[k]}}+\frac{P^{[k]}}{d^{[k]}}\bar{\textbf{H}}^{[kk]}\bar{\textbf{H}}^{[kk]H}\right|
\end{equation}
where $\bar{\textbf{H}}^{[kk]} =\textbf{U}^{[k]H}\textbf{H}^{[kk]}\textbf{V}^{[k]}$

Now, the sum rate achieved over the interference channel is the sum of the rates achieved by all the users $\sum_{k=1}^{K}R^{[k]}$

\subsubsection{Chordal Distance}
\label{chordResults}
It is desirable to keep the signal and interference subspace roughly orthogonal, as interference suppression leads to the loss of the signal component lying in the interference space. This suppression reduces the projection of the desired signal in the interference space resulting in higher sum capacity. Channel realizations corresponding to different states of the antenna, results in varying degree of distance between the interference and signal space. We, therefore, use \textit{chordal distance} (\ref{chordal}), defined over the Grassmann manifold $\mathcal{G}(1,2)$~\cite{love2005limited}, as the distance metric to quantify performance gains: 

\begin{equation}
\label{chordal}
d(\textbf{X},\textbf{Y}) =\sqrt{ \frac{c_{X} + c_{Y}}{2} - \left\| O(\textbf{X})^{\dagger} O(\textbf{Y})\right\|_{F}^{2}},
\end{equation}

where $c_{X}$ denotes the number of columns in matrix $\textbf{X}$ and $O(\textbf{X})$ denotes the orthonormal basis of $\textbf{X}$. The sum rate performance~(\ref{rate}) then becomes a function of the chordal distance between the two spaces~\cite{sung2010linear}. 

\begin{align}
\label{netchordal}
\begin{split}
D = d(\textbf{H}^{\left[1,1\right]}\textbf{v}^{\left[1\right]},\textbf{H}^{\left[1,2\right]}\textbf{v}^{\left[2\right]})
%
\end{split}
\end{align}

%

\section{Simulation Setup and Performance Analysis}
\label{perfAnalysis}

We simulate a MIMO interference network employing spatial IA using conventional technique and present results for $K=3$ users and DoF $d^{[k]}=1$. Further, each node is equipped with $M=2$ antennas. The antennas at the transmitter are conventional omnidirectonal antennas and the antennas at the receivers are reconfigurable antennas with $\mathcal{P}$ states. We provide results for $\mathcal{P}=4$. Since each receiver can select from $\mathcal{P}=4$ states, the cardinality $C$ of the set $I$ of the combinations for $3$ receivers is $\mathcal{P}^K=64$. In other words, the bandit algorithms described above selects from $C=64$, $K$-dimensional vectors.  We also note that, in practice, the proposed sequential learning framework is neither specific to any IA technique nor to specific type of reconfigurable antennas. 


We evaluate the performance of the proposed online bandit algorithms using the simulation setup described above. We compare the proposed algorithms against three selected policies: 1) Oracle Policy: In this policy, it is assumed that there is an oracle which knows the true mean rewards associated with all the antenna state combinations \textit{apriori} and always selects the optimal antenna state combination $c*$ at every time slot. This oracle closely represents the ideal case where instantaneous full CSI corresponding to all the antenna state combinations are available to all the receivers and a centralized controller can exhaustively select for the optimal antenna state combination. 2)  Conventional IA: This technique is a conventional IA scheme which does not employ any reconfigurable antennas at the receiver. Therefore this technique serves as the baseline for the case with conventional antennas and the network does not have means to leverage pattern diversity and there is no need for state selection 3) Random Selection: In this selection scheme, at each time slot,  an antenna state combination vector $c$ is randomly selected with uniform probability from the available set $I$. Further, we generated 1000 channel samples for the distributions associated with each antenna state. For generating the results, we ran the proposed algorithms and other policies 100 times and averaged the result.

\subsection{Regret Analysis}
\begin{figure}[ht]
	\centering
	\includegraphics[width=0.5\textwidth, height=0.4\textwidth]{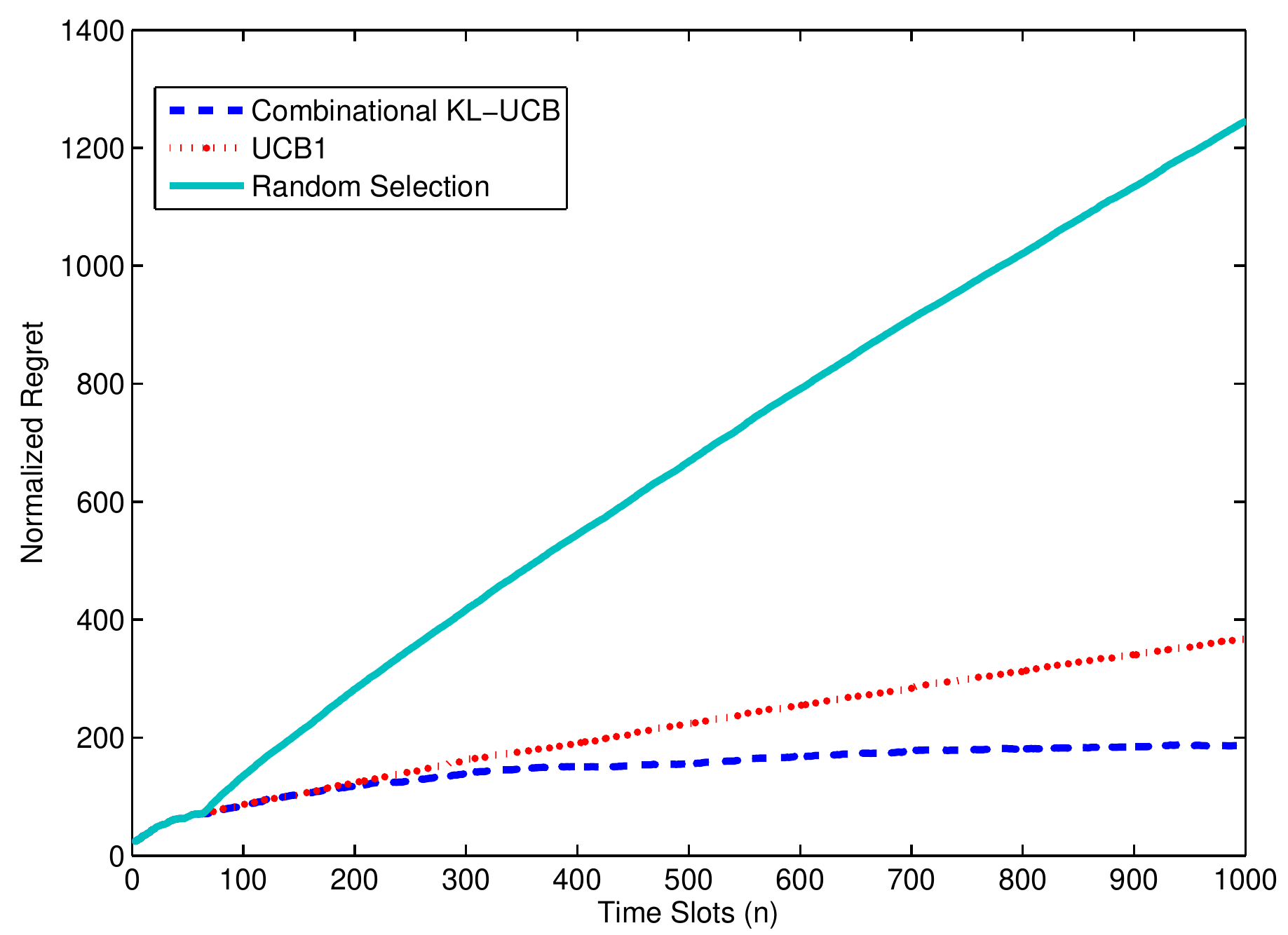}
	\caption{Normalized Regret vs Time Slots (n). $\mathcal{P}=4$}
	\label{fig:regret}
\end{figure}

As described above, one of the ways to evaluate the performance of a multi-armed bandit based sequential algorithm is to calculate the regret. Even though the goal is to maximize long-term performance, regret is a finer performance criteria. In Fig.~\ref{fig:regret}, we show the calculated regret based on Eq.~\ref{laiRegret} with respect to the time slot or decision period $n$ for all the selection policies. It can be clearly seen that the results follow an expected trend. The KL-UCB algorithm has the lower regret followed by UCB1 algorithm. Regret for both the bandit policies is sublinear showing excellent performance. More interestingly, as expected, regret for KL-UCB flattens out and KL-UCB converges much faster and with lower bounded value of regret than UCB1. 

This performance is the result of the KL-UCB algorithm's ability to control the exploration of sub optimal antenna state combination vector faster and increasingly select more optimal choices. On the other hand, random selection policy has almost linear regret with respect to time showing its sub optimality. This poor performance is because the random policy samples all the arms (antenna state combination vectors) with equal probability.

\begin{figure}[ht]
	\centering
	\includegraphics[width=0.5\textwidth, height=0.4\textwidth]{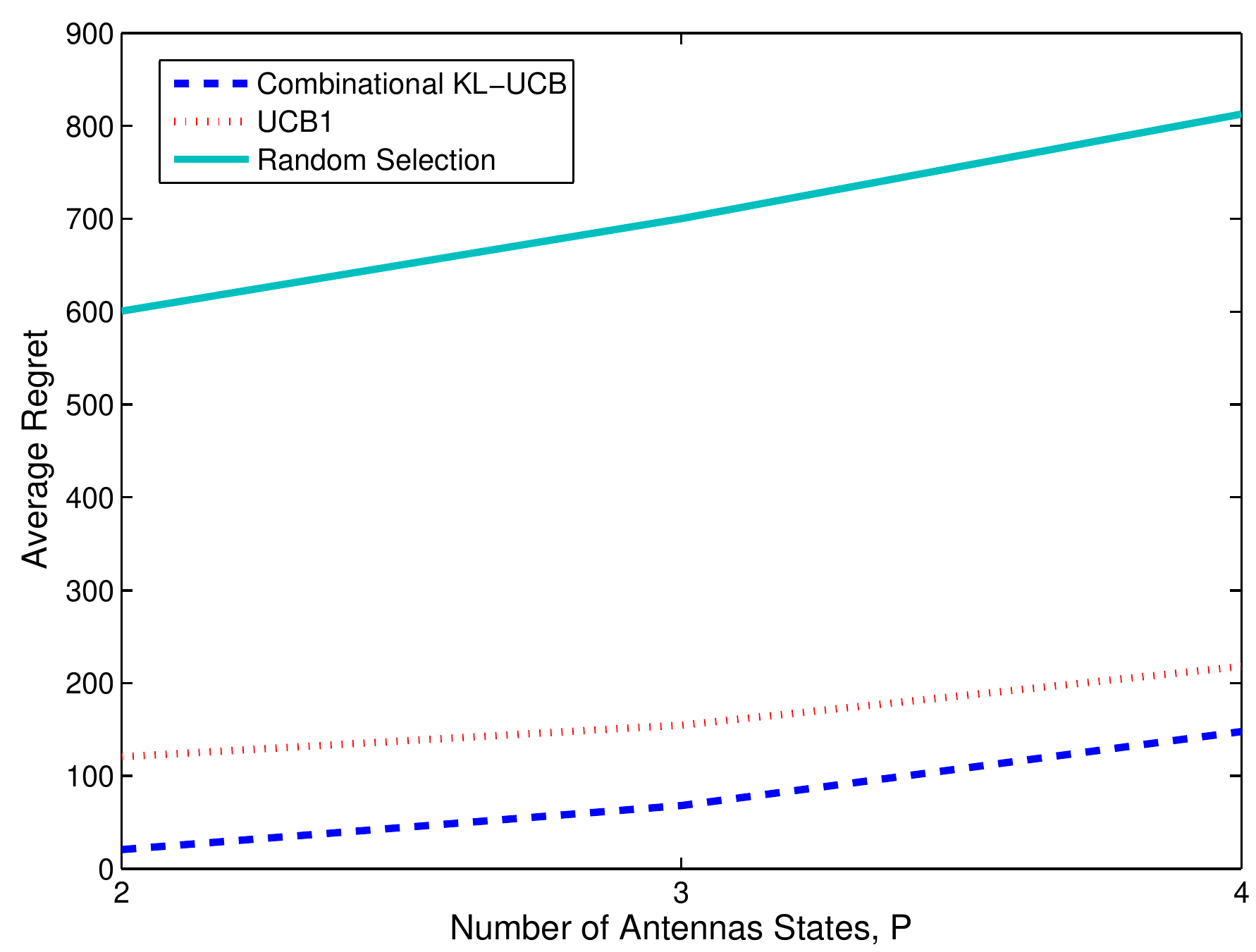}
	\caption{Average Regret vs Number of Antenna States $\mathcal{P}$}
	\label{fig:regretvsstates}
\end{figure}

In Fig~\ref{fig:regretvsstates}, we show how the performance of the proposed algorithms scale with the number of antenna states $\mathcal{P}$ available at each receiver. When $\mathcal{P}$ is increased at each receiver, the total number of combinations vectors $\mathcal{C}$ also increase. The graphs shows that when the antenna states are increased from $\mathcal{P}=2$ to $\mathcal{P}=4$, both UCB algorithms maintain their stability. Note that when $\mathcal{P}=2$ and $\mathcal{P}=4$, $\mathcal{C}=8$ and $\mathcal{C}=64$ respectively. Also, as expected, the average regret for the random policy grows at a faster rate.  

\subsection{Sum Rate Performance}
\begin{figure}[h]
	\centering
	\includegraphics[width=0.5\textwidth, height=0.4\textwidth]{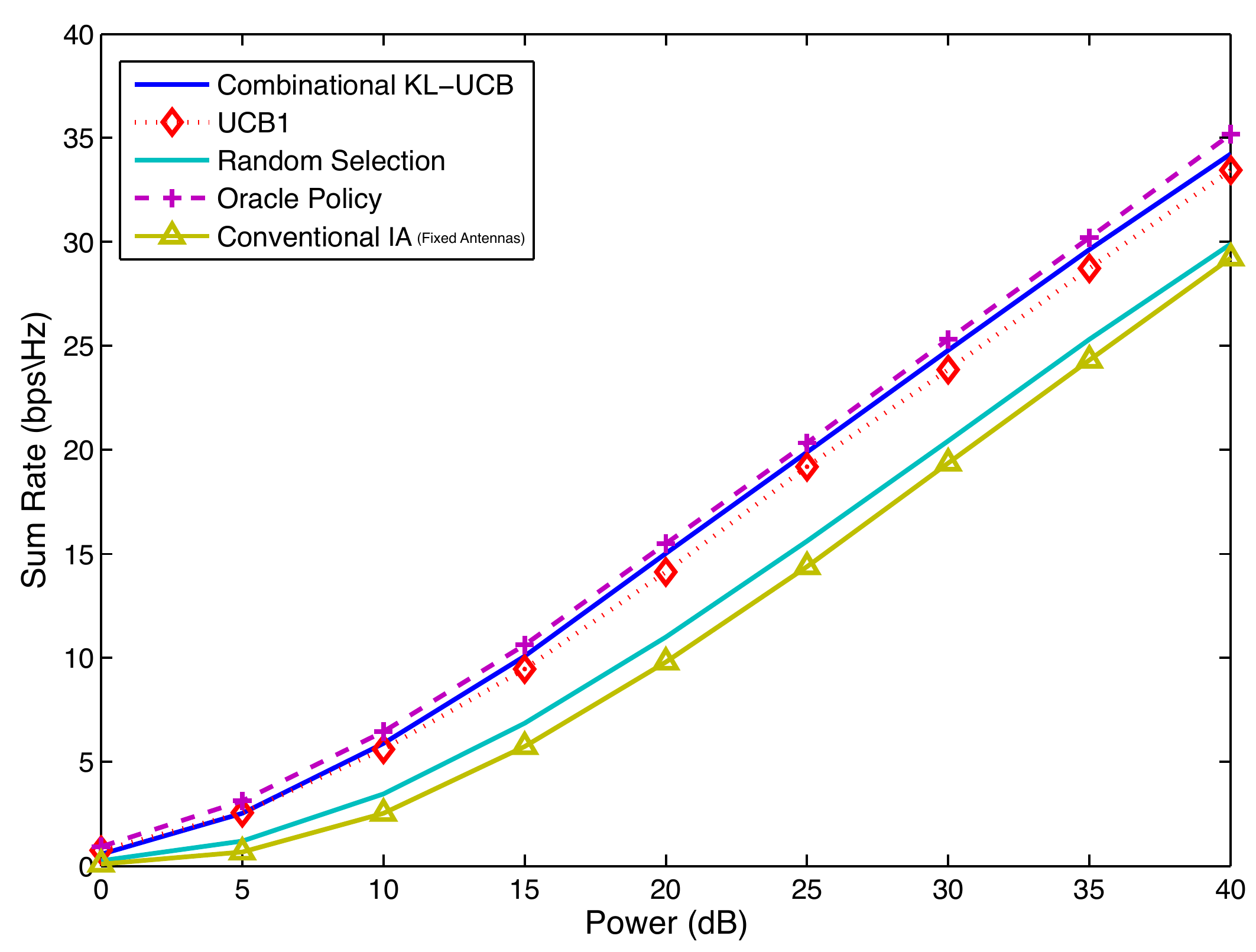}
	\caption{Test:Sum rate vs Power (dB). $\mathcal{P}=4,C=64$, Reward Function-Sum Rate~(\ref{rate})}
	\label{fig:sumrate}
\end{figure}

Lower regret for an given policy should translate into higher average reward. We will now show the performance of the proposed algorithms in terms of the sum rate of the network. In Fig.~\ref{fig:sumrate}, we show the sum rate of the network vs transit power for $K=3$. The baseline comparison with conventional IA scheme with no reconfigurable antennas clearly shows that the algorithms utilizing diversity offered by these antennas are superior in performance. The figure shows that the KL-UCB algorithm performs very close to the oracle policy which is expected. In most non-trivial cases, a learning algorithm will always perform just below the optimal case. Overall, we see nearly $24\%$ percent improvement in sum rate. Another significant result is that sum rate performance is better even in the low and mid SNR regimes. Further, a surprising result is that even the random policy works marginally better than conventional scheme with no reconfigurable antennas. Now, if there was no state selection scheme, the conventional IA scheme in this case will use a fixed antenna state. In the worst case scenario, the receivers will always select the worst channel and in the best case scenario, it will select the best case scenario or any of the other states. But the random selection scheme will at least select all the states with equal likelihood. So, this may cause the random selection to be marginally better than a fixed scheme
  
\subsection{Subspace Distance}
\begin{figure}[h]
	\centering
	\includegraphics[width=0.5\textwidth, height=0.4\textwidth]{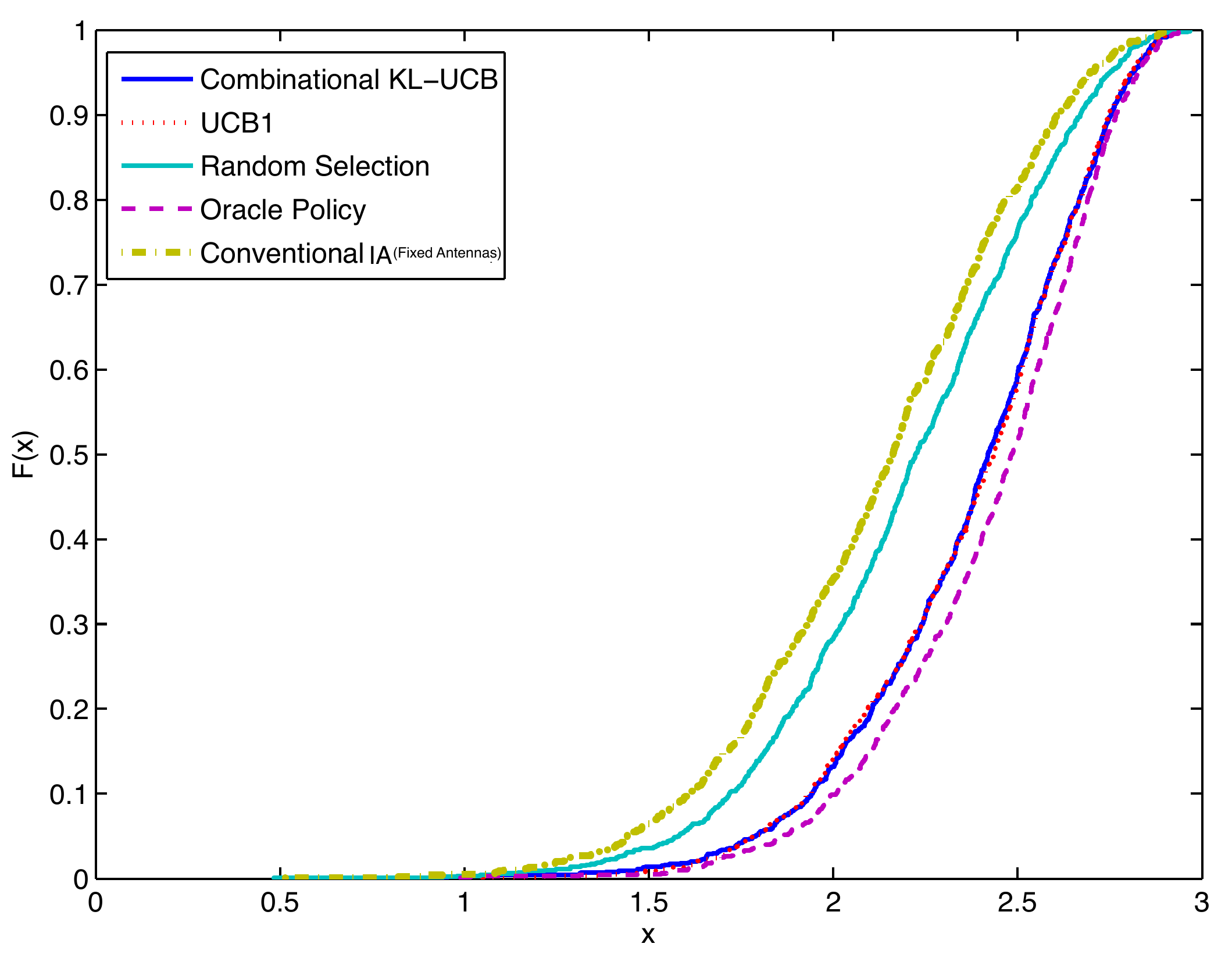}
	\caption{CDF of Chordal Distance. $\mathcal{P}=4$}
	\label{fig:chorddist}
\end{figure}

As mentioned in Sec.~\ref{rewardMetrics}, the idea of using reconfigurable antenna based channel selection for enhanced sum rate for IA comes from the idea of maximizing distance between signal and interference subspaces. In Fig.~\ref{fig:chorddist}, we show the CDF of the total chordal distance as calculated in Eq.~\ref{netchordal}. The figure clearly shows that the maximum chordal distance is achieved by the oracle policy. This performance is closely followed by KL-UCB and UCB1 policy. On the other hand, conventional IA achieves the lowest chordal distance of all the policies consistent with previous results. 

Over all we see that for the performance metrics the trends are similar and the proposed learning policies perform better than the conventional IA scheme.

\begin{figure}
	\centering
	\includegraphics[width=0.5\textwidth, height=0.4\textwidth]{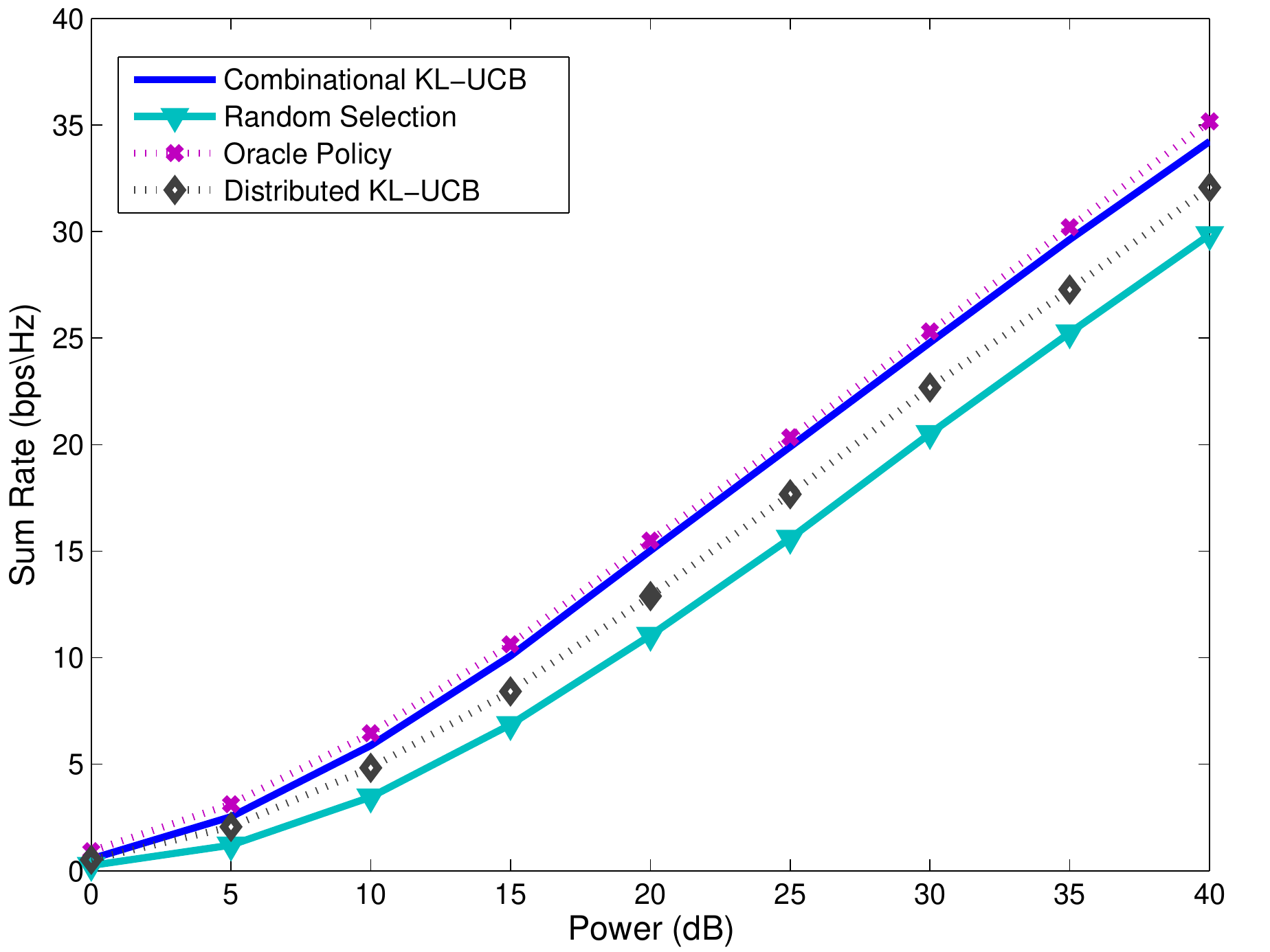}
	\caption{Sum rate vs Power (dB). $\mathcal{P}=4$, Reward Function - Chordal Distance~(\ref{netchordal})}
	\label{fig:sumratechordal}
\end{figure}

\subsection{Distributed Selection with Chordal Distance}
In this section, we show the sum rate results when using chordal distance described above as a reward metric. In the combinational KL-UCB case, the reward is a joint sum rate which is ingested by the bandit controller. In case of the chordal distance, each receiver made its own state selection with its own bandit controller. This naturally means that this technique will not explore all the combinations of antenna states at the receivers. Also, note that for each of the bandit controller now only has $\mathcal{P}$ antenna states to choose from, i.e.,  number of arms is $\mathcal{P}$. In Fig~\ref{fig:sumratechordal}, we show the sum rate achieved for both the combinational and distributed version of KL-UCB algorithm alongwith the Oracle and random policy. For clarity we will drop the comparison with other schemes which perform worse. Fig~\ref{fig:sumratechordal}, shows that the combinational version still outperforms the other policies. This is expected as the distributed KL-UCB works in somewhat greedy fashion where each node only maximizes its own chordal distance. In other words, it only selects the channel which produces lowest SNR loss at that receiver. On the other hand, in the combinational version,  a joint network utility, i.e,. the sum rate is maximized. On the other hand, distributed KL-UCB still performs better than the random policy.
  
\section{Conclusion}
\label{conclusion}

In this paper we have shown that optimal channel selection for interference alignment can significantly improve performance of the users in the network. We have proposed practical sequential algorithms to amortize the complexity cost and adaptively select between a combination of antenna states at the receivers. Through extensive simulations we have shown that that sum rate can be maximized and the channel selection via reconfigurable antennas add another degree of freedom to optimize IA performance even in low and mid SNR regimes.

The results from this paper can be used as a motivation and starting point for further research in understanding the benefits of reconfigurable antennas based diversity for interference alignment. Performance analysis using a MIMO IA testbed, incorporating radiation state correlation and its impact on bandit policy design are interesting areas of future work. 


\section*{Acknowledgment}
This material is based upon work supported by the National Science Foundation under Grant No. 1422964.


\bibliographystyle{IEEEtran}
\bibliography{IEEEabrv,Globecom2012Example,APS_Special_Issue_MAB_Bibliography,RWSExample}

\end{document}